\documentclass[aps,floats,floatfix,amssymb,amsmath,prb,nofootinbib,twocolumn,superscriptaddress]{revtex4-2}

\usepackage{physics} 
\usepackage[T1]{fontenc}
\usepackage{graphicx}
\usepackage{hyperref}
\usepackage{babel}
\usepackage{amsfonts}
\usepackage{color}
\usepackage{capt-of}
\usepackage{multirow}
\usepackage{xspace}
\usepackage[normalem]{ulem}

\begin{document}
\title{Quantum-classical study of charge transport in organic semiconductors with multiple low-frequency vibrational modes}

\author{Darko Tanaskovi\'c}
%\email{tanasko@ipb.ac.rs}
\affiliation{Institute of Physics Belgrade, University of Belgrade, 
Pregrevica 118, 11080 Belgrade, Serbia}

\author{Maksim Makrushin}
\affiliation{Institute of Physics Belgrade, University of Belgrade, 
Pregrevica 118, 11080 Belgrade, Serbia}

\author{Petar Mitri\'c}
%\email{mitricp@ipb.ac.rs}
\affiliation{Institute of Physics Belgrade, University of Belgrade, 
Pregrevica 118, 11080 Belgrade, Serbia}

\begin{abstract}
Building on the recent success of a quantum–classical method for computing transport properties in the Holstein model with a single phonon mode [P. Mitri\'c {\it et al.}, Phys. Rev. B {\textbf{111}}, L161105 (2025)], we now assess its reliability in more realistic scenarios involving multiple phonon modes in the Holstein model, as well as single- and multi-mode Peierls models. For parameters relevant to the prototypical organic semiconductor rubrene, we compute the frequency-dependent charge mobility and find excellent agreement with results from the state-of-the-art hierarchical equations of motion method. These results show that the method, previously validated only for the single-mode Holstein model, preserves quantitative accuracy in substantially more complex and material-relevant regimes. Our microscopic approach complements the phenomenological transient-localization theory and is readily applicable to realistic electron–phonon Hamiltonians.
\end{abstract}

\maketitle

%%%%%%%%%%%%%%%%%%%%%%%% INTRODUCTION %%%%%%%%%%%%%%%%%%%%%%%%%%%%%%%%%%%%%%
%{\it Introduction.}
\section{Introduction}
Charge transport in organic semiconductors (OSCs) is primarily limited by electron scattering on thermally excited low-frequency vibrational modes \cite{Coropceanu2007,Zhugayevych2015,Fratini2020,Nematiaram_2020}. 
For these systems, the simple quasiparticle picture underlying the semiclassical Boltzmann approach breaks down, calling for alternative theoretical descriptions \cite{2015_DeFilippis, 2020_Li, 2021_Bertini, 2024_Ostmeyer, Buividovich:2024MY}. A major milestone was the transient localization (TL) approach introduced by Fratini {\it et al.}~\cite{ Fratini2016,Ciuchi_PRB2011}, who observed that at short times the electron experiences an almost static but randomly distributed potential, revealing precursors of Anderson localization, whereas at longer times lattice dynamic suppresses interference and restores finite dc conductivity even in one dimension (1D). Within the phenomenological TL framework, this behavior is captured by first evaluating the current–current correlation function of the fully static Anderson model and then introducing an exponential damping to account for the fact that the disorder is dynamic rather than static. This simple approach, whose hallmark is the displaced Drude peak in optical conductivity, forms the basis of our current understanding and provides a remarkably accurate description of transport in OSCs~\cite{Fratini_NatMat2017}.

%\vspace*{-.1cm}

A more microscopic route was proposed by Troisi and Orlandi \cite{Troisi_PRL2006}, who treated soft vibrational modes classically, while the electron dynamics was obtained from the time-dependent Schrödinger equation; the charge mobility was then obtained using linear-response theory. This seemingly straightforward approach to computing dc mobility was later questioned \cite{Fratini2016,Nematiaram_2020}, as long-time diffusion was thought to be unreliable due to the unphysical electron energy increase with propagation time—a well-known artifact of quantum–classical (QC) dynamics \cite{Parandekar_2005,Tully_2023}. This view has recently begun to shift \cite{Fetherolf_PRX2020,Runeson_2024,Mitric_Precursors_2025}. In particular, in Ref.~\cite{Mitric_Precursors_2025} we showed, using the one-dimensional Holstein model with a single phonon mode, that QC mobility agrees remarkably well with the fully quantum result. In the weak-to-intermediate coupling regime, the current–current correlation function was found to decay to zero before the electronic kinetic energy changed significantly, explaining the success of the QC approach. Within regimes where TL is expected to be accurate, QC and TL optical conductivities, $\mu(\omega)$, were in close agreement—except for the zero-frequency peak, present in the QC and fully quantum results, but absent in TL. Recent work has shown that when multiple phonon modes are included, this zero-frequency peak can be  suppressed in the fully quantum solution, making the TL prediction for dc conductivity more accurate \cite{Jankovic_DEOM_2025}. However, it remains to establish the applicability of the QC approach to a more realistic case of the off-diagonal dynamic disorder and multiple vibrational modes. This is particularly important since one of the key questions that has emerged in a quest for the synthesis of high mobility OSCs is whether a single or multiple modes
dictate charge transport \cite{Peluzo_2025,Schweicher_2019}.

%\vspace*{-1.5cm}
In this work, we assess the applicability of the QC method in the presence of multiple low-frequency phonon modes. Using the Holstein model, we show that multiple phonon modes, within QC, can entirely suppress the zero-frequency peak in $\mu(\omega)$, thereby bringing the QC results into closer agreement with the TL prediction. We then turn to the 1D Peierls model with parameters relevant to rubrene at room temperature, where we find excellent agreement with state-of-the-art numerically “exact’’ hierarchical equations of motion calculations \cite{Jankovic_HEOMII_2025,Jankovic_DEOM_2025}. Analogously to the Holstein model, the zero-frequency peak \cite{Runeson_2024} gets also suppressed in the Peierls model with multiple low-frequency modes. These results demonstrate that the QC method offers a simple and computationally efficient framework for evaluating charge mobility with realistic phonon spectra, providing a promising route to simulations of charge transport in higher-dimensional and material-specific systems.

%{\it Model and methods.}
\section{Model and methods} 
%\vspace*{-.2cm}
The 1D Hamiltonian that contains both the inter-site (Peierls) coupling, which modulates the electron hopping, and the on-site (Holstein) term is given by
\begin{align} \label{eq:Holstein_hamiltonian}
H =& -t_0 \sum_i  \left( c_i^\dagger c_{i+1} + \mathrm{H.c.} \right)   \nonumber\\
   & + \sum_{i \xi} g_\xi^{\mathrm{P}}  \sqrt{2\omega_\xi}(x_{i\xi} - x_{i+1,\xi}) \left( c_i^\dagger c_{i+1} + \mathrm{H.c.} \right)  \nonumber\\
   &  +\sum_{i \xi} g_\xi^{\mathrm{H}}  \sqrt{2\omega_\xi} x_{i \xi} c_i^\dagger c_i + \sum_{i\xi} \omega_\xi a_{i\xi}^\dagger a_{i\xi}.
\end{align}
Here, $t_0$ is the hopping parameter (transfer integral), $c_i^\dagger$ ($a_{i\xi}^\dagger$) is the electron (phonon) creation operator, and we assume that there is only a single electron in the band, as appropriate for  weakly doped semiconductors. $x_{i\xi}$ is proportional to the displacement operator for vibrational mode $\xi$, $x_{i\xi} = (1/\sqrt{2 \omega_\xi}) (a_{i\xi}^\dagger + a_{i\xi})$.
The electron coupling to different vibrational modes of frequency $\omega_\xi$ is determined by the coupling constants $g_\xi^{\mathrm{P}}$ and $g_\xi^{\mathrm{H}}$. $t_0$, $\hbar$, $k_B$, $e$ and the lattice constant are set to one. To examine the role of multiple phonon modes, we concentrate on purely Holstein $(g_\xi^{\mathrm{P}} = 0)$ or purely Peierls type of coupling $(g_\xi^{\mathrm{H}} = 0)$. For the Holstein model we make a comparison with the single-mode case for a parameter set that was studied in exceptional detail in a series of recent papers \cite{Mitric_Precursors_2025,Mitric_QT_2025,Jankovic_2024,Jankovic_2023}. For the Peierls case, we use the parameters that are standard in the studies of quasi-1D rubrene \cite{Fratini2016,Fetherolf_PRX2020,Runeson_2024}. In both cases, a somewhat realistic choice of $g_\xi$ is obtained when they are selected to mimic the Brownian-oscillator spectral density \cite{Jankovic_DEOM_2025}
\begin{equation}\label{eq:BO}
g_{\xi}^2  = \alpha \frac{2\gamma_0  \omega_0^2 \omega_\xi}{(\omega_0^2-\omega_\xi^2)^2 + (2\gamma_0 \omega_\xi)^2},
\end{equation}
while for $\omega_{\xi}$, in practice, we use $11$ vibrational modes uniformly distributed in the range $(0.1\omega_0,1.9\omega_0)$. In Eq.~\eqref{eq:BO}, $\gamma_0$ is the damping parameter, while $\alpha$ is a proportionality constant (independent of $\xi$) that will be defined later [see Eq.~\eqref{eq:prop_const_alfa}]. In the limit $\gamma_0 \rightarrow 0$, only a single vibrational mode, $\omega_\xi = \omega_0$, has nonzero coupling $g_\xi$, reducing the system to a single-mode case. The corresponding coupling is arbitrary and will be denoted $g_0$ throughout this work. Dimensionless coupling strengths are then typically defined as $\lambda^\mathrm{H} = g_0^2/(2\omega_0 t_0)$ and $\lambda^\mathrm{P} = 2 g_0^2/(\omega_0 t_0)$ for the Holstein and Peierls cases, respectively.

We will solve the model using the QC method which treats the ion dynamics classically and solves the time-dependent electron Schr\" odinger equation \cite{Troisi_PRL2006,Mitric_Precursors_2025}. The QC Hamiltonian is obtained by replacing the operator  $x_{i\xi}$ by a classical variable. The electron Hamiltonian in the Holstein case is thus given by
\begin{equation}\label{eq:QC_Holstein}
 H^{\mathrm{H}}_{\mathrm{el}} = -t_0 \sum_i \left( c_i^\dagger c_{i+1} + \mathrm{H.c.} \right) + \sum_{i\xi}  g_\xi^{\mathrm{H}} \sqrt{2\omega_\xi} x_{i\xi} c_i^\dagger c_i,
\end{equation}
whereas in the Peierls case 
\begin{align} \label{eq:QC_Peierls}
H^{\mathrm{P}}_{\mathrm{el}}  &= -t_0 \sum_i  \left( c_i^\dagger c_{i+1} + \mathrm{H.c.} \right)   \nonumber \\
   & + \sum_{i \xi} g_\xi^{\mathrm{P}} \sqrt{2\omega_\xi}(x_{i\xi}-x_{i+1,\xi}) \left( c_i^\dagger c_{i+1} + \mathrm{H.c.} \right) .
\end{align}
Hence, the dynamic disorder enters through diagonal (off-diagonal) terms for the Holstein (Peierls) model.
In Ref.~\cite{Mitric_Precursors_2025} we showed that the back-action term (from the electron to ions) has a negligible effect on the electron dynamics. Therefore, we can set harmonic oscillations for each mode, $x_{i\xi}(t) = x_{i\xi}(0)\cos(\omega_\xi t) + (\dot x_{i\xi}(0)/\omega_\xi) \sin(\omega_\xi t)$. The initial displacements and velocities need to be taken from Gaussian distributions with the variance $\langle x_{i\xi}^2 (0) \rangle = \frac{1}{2\omega_\xi} \coth \left( {\beta \omega_\xi} / {2}\right)$ and $\langle \dot x_{i\xi}^2(0) \rangle = \frac{\omega_\xi}{2} \coth \left( {\beta \omega_\xi} / {2}\right)$, respectively (see, e.g., the Supplemental Material of Ref.~\cite{Mitric_Precursors_2025} for a derivation), where $\beta = 1/T$ is the inverse temperature. 
The short-time dynamics depend only on the effective diagonal (off-diagonal)  disorder $\varepsilon_{i} =  \sum_{\xi}  g_\xi^{\mathrm{H}} \sqrt{2\omega_\xi} x_{i\xi}(0)$  ($\delta t_{0i,i+1} = \sum_{\xi} g_\xi^{\mathrm{P}} \sqrt{2\omega_\xi}(x_{i\xi}(0)-x_{i+1,\xi}(0))$). This dynamic disorder also obeys the Gaussian distribution, now with the variance 
$\langle \varepsilon_i^2 \rangle = \sum_\xi (g_\xi^{\mathrm{H}})^2 \coth  ({\beta \omega_\xi} / {2})$ for the Holstein model and 
$\langle \delta t_{0i,i+1}^2 \rangle = 2 \sum_\xi (g_\xi^{\mathrm{P}})^2 \coth  ({\beta \omega_\xi} / {2})$ for the Peierls model. Since our goal is to examine how the long-time electron dynamics changes when different $\gamma_0$ are selected, it is convenient to keep the short-time dynamics the same as in the $\gamma_0 \to 0$ case.  This is the condition from which the proportionality constant $\alpha$, from Eq.~\eqref{eq:BO} can be determined
\begin{equation} \label{eq:prop_const_alfa}
\alpha = \frac{g_0^2 \coth \left( \frac{\beta\omega_0}{2} \right)}{
\sum_\xi 2\gamma_0  \omega_0^2 \omega_\xi \left[
(\omega_0^2-\omega_\xi^2)^2 + (2\gamma_0 \omega_\xi)^2
\right]^{-1}
\coth \left( \frac{\beta\omega_\xi}{2} \right)
}.
\end{equation}
The QC equations $i\frac{\partial}{\partial t} |\psi_n(t)\rangle = H_{\mathrm{el}}(t) |\psi_n(t)\rangle$ are propagated using a fourth-order Runge–Kutta scheme. As initial conditions, we employ the electronic eigenstates $\psi_n(0)$ of $H_{\mathrm{el}}$, $H_{\mathrm{el}}(0) |\psi_n(0)\rangle = E_n |\psi_n(0)\rangle$, computed for a configuration of randomly displaced ions.

We use the Kubo linear-response formalism \cite{Mahan,Kubo_1957,Mitric_Precursors_2025}, computing the current–current correlation function
\begin{equation}
 C_{jj}(t) =  \frac{1}{Z} \sum_{n,m} e^{-\beta E_n} \langle \psi_n(t)| j |\psi_m(t)\rangle \langle \psi_m| j |\psi_n \rangle ,
\end{equation}
which needs to be averaged over many realizations of initial ion positions and velocities. The current operator  $j$ is ${i t_0 \sum_i \left( c_{i+1}^\dagger c_i - c_i^\dagger c_{i+1} \right)}$ for the Holstein model, and ${i  \sum_i \left[ t_0 - \sum_\xi g_\xi^{\mathrm{P}} \sqrt{2\omega_\xi}(x_{i\xi}-x_{i+1,\xi})\right]  \left( c_{i+1}^\dagger c_i - c_i^\dagger c_{i+1} \right)}$ for the Peierls model, while the frequency-dependent mobility reads as
\begin{equation}
   \mu(\omega) = \frac{2 \tanh \left( \frac{\beta \omega}{2} \right)}{\omega}
  \int_0^\infty dt \cos(\omega t) \mathrm{Re} C_{jj}(t) .
\end{equation}
\begin{figure}[b]
\includegraphics[width=\columnwidth]{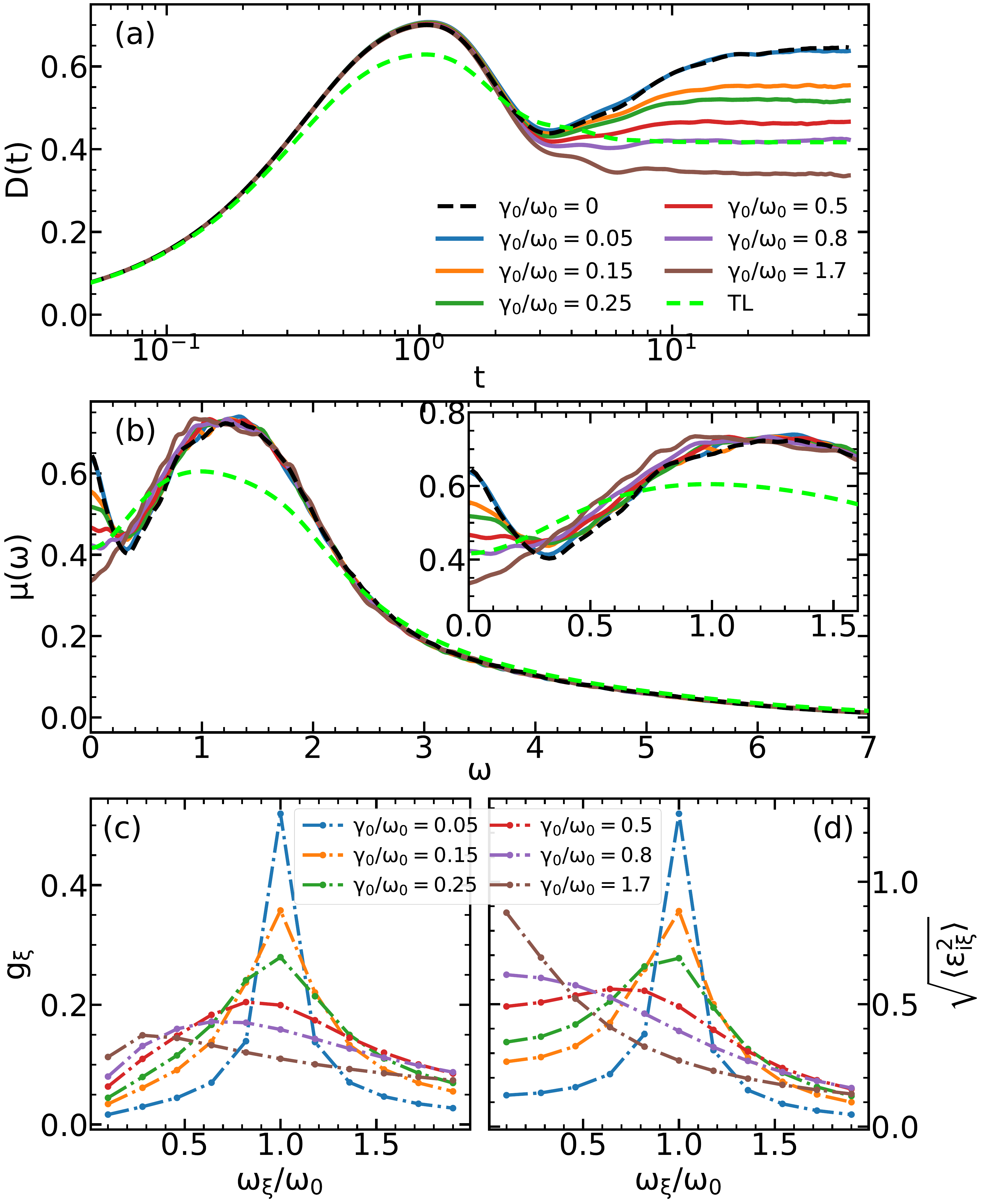}
\caption{(a) Time-dependent diffusion constant and (b) frequency-dependent mobility for (c) several distributions of coupling constants for the Holstein model. (d) The corresponding mode-dependent dynamic disorder standard deviation. Here $T=1$, $\omega_0 = 1/3$ and $\lambda^{\mathrm{H}}=0.5$. 
}
\label{Figure1}
\vspace*{-0.5cm}
\end{figure}
The time-dependent diffusion constant $D(t)$ is equal to  $\int_0^t dt' \mathrm{Re} C_{jj}(t')$ and the dc mobility is given by the Einstein relation ${\mu_{\mathrm{dc}} = D(\infty)/T}$. 

The TL result is obtained by evaluating $C_{jj}(t)$ in the single-mode $\gamma_0 \to 0$ static (Anderson model) limit, $\omega_0 \to 0$, denoted $C_{jj}^\mathrm{AM}(t)$, and by applying an artificial exponential damping, $C_{jj}^\mathrm{TL}(t) = C_{jj}^\mathrm{AM}(t)\, e^{-t \omega_0}$, to capture the dynamic nature of the disorder.

%\clearpage
%\newpage
%\pagebreak

%{\it Results.} 
\section{Results}
%
%{\it A.} 
\subsection{Multiple-mode Holstein model}

We first examine the impact of the multiple vibrational modes in the case of the Holstein model. The parameters are chosen such that in the limit $\gamma_0 \rightarrow 0$ the phonon frequency is $\omega_0 = 1/3$, while the dimensionless coupling constant is $\lambda^{\mathrm{H}} =  0.5$.   The single-mode case was previously studied in many details, using both the numerically exact quantum typicality \cite{Mitric_QT_2025} and hierarchical equations of motion (HEOM) method \cite{Jankovic_2023,Jankovic_2024}, as well as the QC approach \cite{Mitric_Precursors_2025}. The agreement between the results obtained with different methods was excellent, showing that at temperatures $T \gtrsim \omega_0$ the phonon quantization weakly affects charge dynamics. The unexpected result was the upturn in the diffusion constant $D(t)$ at $t \approx 1/\omega_0$ corresponding to the zero-frequency peak in $\mu(\omega)$. The appearance of the zero-frequency peak was not present in the TL scenario \cite{Mitric_Precursors_2025}. 
\begin{figure}[b]
\includegraphics[width=\columnwidth]{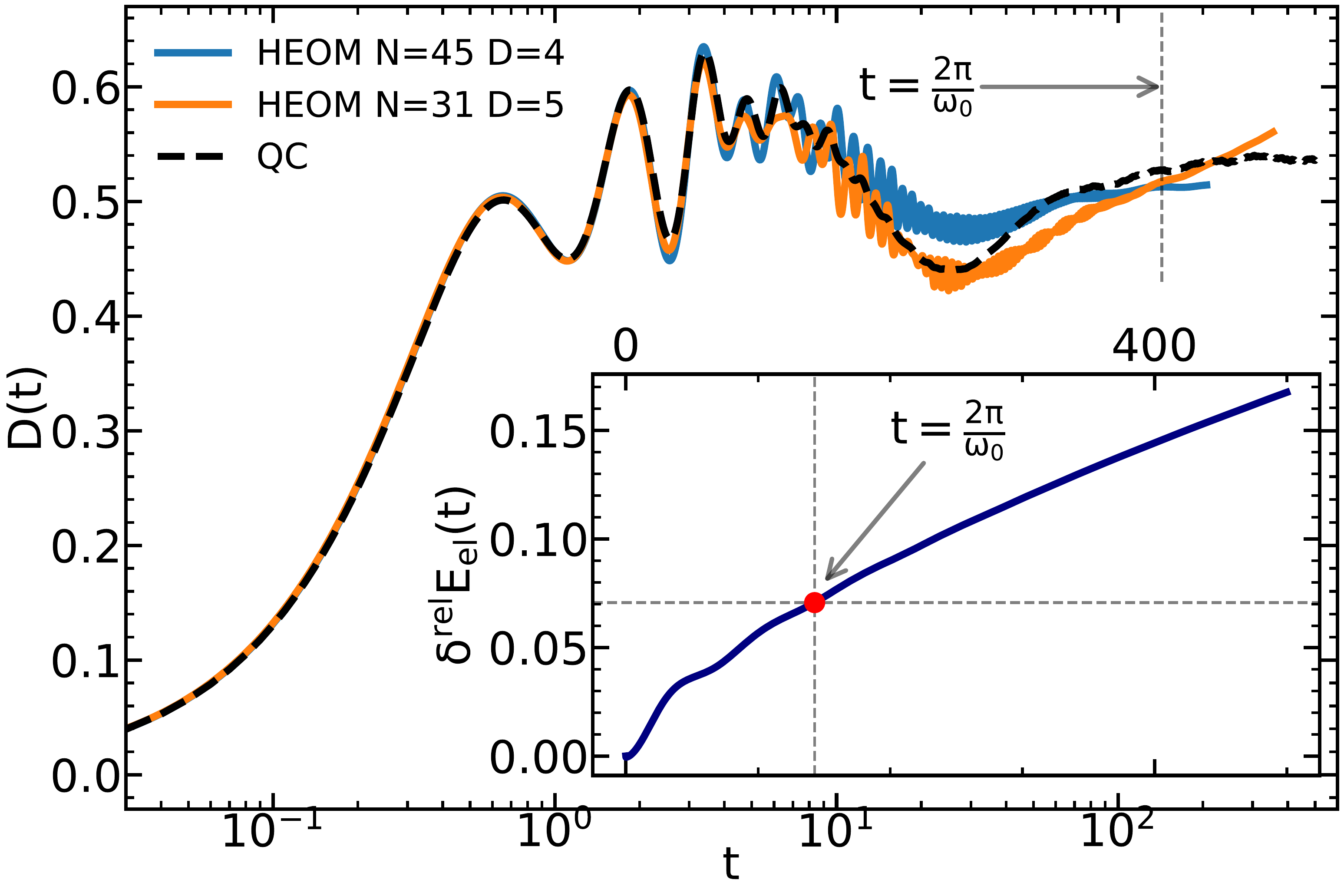}
\caption{Comparison of the QC (dashed line) and HEOM (solid lines) time-dependent diffusion constant for the Peierls model at $T=0.175$, $\omega_0 = 0.044$, $\lambda^{\mathrm{P}}=0.336$. The inset shows the relative increase of the electron kinetic energy with the propagation time in QC equations.
}
\label{Figure2}
\vspace*{-0.5cm}
\end{figure}

The time-dependent diffusion constant $D(t)$ and the frequency-dependent mobility $\mu(\omega)$, for several 
 values of damping $\gamma_0$, are shown in Figs.~\ref{Figure1}(a) and (b), respectively. $g_\xi^{\mathrm{H}}$ and the corresponding mode-dependent standard deviation of diagonal disorder, $\sqrt{\langle \varepsilon_{i\xi}^2 \rangle} = g_\xi^{\mathrm{H}} \left[  \coth  ({\beta \omega_\xi} / {2}) \right] ^{1/2}$, are shown in Figs.~\ref{Figure1} (c) and (d). Since we keep the total disorder variance $\langle \varepsilon_i^2 \rangle$ fixed, the short-time dynamics remains the same for all $\gamma_0$. For small $\gamma_0$, the electron scattering from the central mode of frequency $\omega_0=1/3$ is dominant.
As we increase $\gamma_0$, the other vibrational modes also influence the electron dynamics. In particular, for large values of $\gamma_0$  the electron scattering from low-frequency modes, which experience stronger thermal fluctuations, becomes more important. The increase of $\gamma_0$ is followed by weakening and, eventually, the disappearance of the upturn in $\sigma(\omega)$ for $\omega<\omega_0$, making the charge dynamics closer to the TL prediction \cite{Fratini_PRR2020,Jankovic_DEOM_2025}.

%{\it B.} 
\subsection{Single-mode Peierls model}
We benchmark the QC with the fully quantum HEOM solution \cite{Jankovic_HEOMII_2025} for the single vibrational mode 1D Peierls model in Fig.~\ref{Figure2}. The parameters $\omega_0=0.044$, $T=0.175$ and $\lambda^{\mathrm{P}} = 0.336$ are typically used for a description of room-temperature rubrene \cite{Troisi_2007,Girlando_2010,Ordejon_2017,Fratini2016,Fetherolf_PRX2020}. We note that in physical units $t_0 = 143 \, \mathrm{meV}$ and $\omega_0 = 6.2 \,\mathrm{meV}$. The HEOM solution that we take from Ref.~\cite{Jankovic_Zenodo_Peierls}, is in principle exact. This quite complex and numerically demanding method  has two key numerical parameters: the length of the 1D lattice $N$ and the hierarchy depth $D$. For the QC method we used $N=200$ (much shorter chain gives very similar result) and averaged the correlation function over 10000 realizations of initial displacements and velocities. We ascribe the small discrepancy between the QC and HEOM solutions to the small uncertainty in HEOM results which are not fully converged with respect to $N$ or $D$. For $t \sim 20$ the QC result is closer to the HEOM $D=5, N=31$ solution, but at longer times it becomes closer to the  $D=4$ solution on a longer $N=45$ lattice. We note that regime parameters correspond to a low phonon-frequency regime ($T/\omega_0 \approx 4$) and weak electron–phonon coupling, where--consistent with our Holstein-model results \cite{Mitric_Precursors_2025}--the QC method is expected to accurately capture the quantum dynamics. Moreover, as discussed in the introduction, the time evolution of the electron kinetic energy further indicates that the QC approximation should remain reliable in our case. In particular, although the ensemble-averaged electron kinetic energy $\langle E_{\mathrm{el}}(t) \rangle$ increases with time and eventually approaches the band center, $\langle E_{\mathrm{el}}(t \to \infty) \rangle = 0$, the inset of Fig.~\ref{Figure2} shows that it changes only modestly by the time the diffusion coefficient $D(t)$ reaches a well-defined plateau at $t_\mathrm{plateau} \approx 2\pi/\omega_0$. This is quantified by $\delta^{\mathrm{rel}}E_{\mathrm{el}} (t)= 1- |\langle E_{\mathrm{el}}(t) \rangle / \langle E_{\mathrm{el}}(0) \rangle |$, which remains small at that point, $\delta^{\mathrm{rel}}E_{\mathrm{el}}(t_\mathrm{plateau}) \approx 0.07$ (note that $\delta^{\mathrm{rel}}E_{\mathrm{el}}(t\to\infty) = 1$). This gives us additional confidence to the precision of the QC method. Let us also note that we have further verified that the back-action term makes a negligible contribution to the electron dynamics, consistent with previous studies \cite{Fetherolf_PRX2020,Runeson_2024,Mitric_Precursors_2025}.

\begin{figure}[t]
\includegraphics[width=\columnwidth]{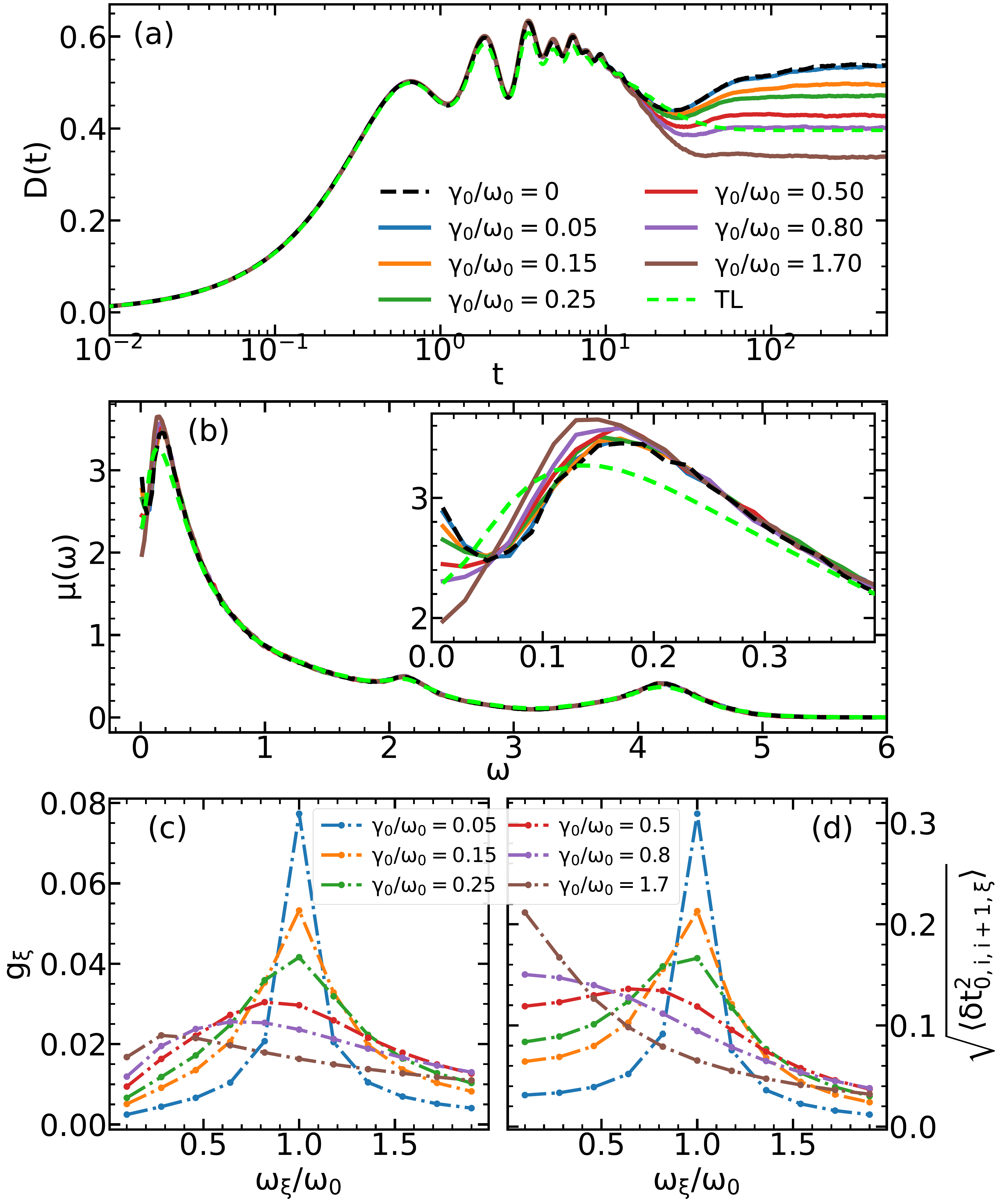}
\caption{(a) Time-dependent diffusion constant and (b) frequency-dependent mobility for (c) several distributions of coupling constants for the Peierls model. (d) The corresponding mode-dependent dynamic disorder standard deviation. Here $T=0.175$, $\omega_0 = 0.044$ and $\lambda^{\mathrm{P}}=0.336$. 
}
\label{Figure3}
\vspace*{-0.5cm}
\end{figure}

%{\it C.} 
\subsection{Multiple-mode Peierls model}
The impact of multiple vibrational modes to diffusion and mobility in the Peierls model is shown in Figs.~\ref{Figure3} (a) and (b). In the $\gamma_0 \rightarrow 0$ limit the model reduces to the single-mode case that we have just described. We take the same set of phonon frequencies ratio $\omega_\xi/\omega_0$ as for the Holstein model, while the distribution of coupling constants and mode-dependent off-diagonal disorder is shown in Figs.~\ref{Figure3} (c) and (d). For larger $\gamma_0$, the scattering from the modes away from $\omega_0$ becomes more important. This leads to the gradual disappearance of the upturn in $D(t)$ and the zero-frequency peak in $\mu(\omega)$. The TL solution is shown for comparison. Interestingly, the zero-frequency peak was observed in some experiments on OSCs \cite{Fischer_2006}, while it is absent in others \cite{Li_2007}.

Our QC solution is in quite good agreement with a very recent dissipation equations of motion (DEOM) solution shown in Fig.~5 of Ref.~\cite{Jankovic_DEOM_2025}. We believe that a discrepancy is mostly due to the insufficient hierarchical depth $D = 4$ in the DEOM solution, which particularly affects the low-frequency modes due to low phonon excitation energy (see a comparison in our Fig. 2 and a discussion in Sec. III of Ref.~\cite{Jankovic_DEOM_2025}).

%{\it Conclusions.}
\section{Conclusions}
In summary, we showed that the QC solution for frequency-dependent mobility in the 1D Peierls model with parameters representative of rubrene is in a very good agreement with state-of-the-art numerically ``exact'' hierarchical equations of motion solution \cite{Jankovic_HEOMII_2025,Jankovic_DEOM_2025}. Together with the previous results on the Holstein model \cite{Mitric_Precursors_2025}, our work shows that the QC method gives an excellent approximation for the charge transport at temperatures $T \gtrsim \omega_0$ and weak-to-intermediate electron-phonon coupling. These simple conclusions remove a misconception that a QC approach cannot be used as a reliable tool for the calculation of dc mobility due to the artificial electron heating \cite{Nematiaram_2020,Fratini2016}, which withheld a wider use of the QC calculations in the literature. As we here demonstrate on the example of rubrene parameters, a significant electron heating appears only after the plateau in $D(t)$ is reached and $C_{jj}(t) \approx 0$, which explains the success of the QC approach. This no longer holds in the regime of strong electron-phonon coupling, when the QC result cannot be used as a good prediction of the dc mobility. Nevertheless, for weak-to-intermediate couplings, where QC works well, the calculation can be easily done for an arbitrary distribution of low-frequency vibrational modes and coupling constants, which can be of interest for recent applications of the QC approach in different physical contexts \cite{Fetherolf_2023,Wang_2011,Brink_2022,Menzler_2025,Mayers_2018,Nguyen_2025,Aydin_PRL2024,Aydin_2024,Krotz_2024}.
For a narrow distribution of coupling constants around the dominant one, a central peak appears in the frequency-dependent mobility, which gives a prediction for the dc mobility few tens of percent higher than in the TL scenario. This central peak disappears for a broad distribution of vibrational modes and coupling constants. 

The application of the QC approach in two dimensions
should be as simple and numerically feasible as we have
here in 1D, which is particularly important for layered systems such as OSCs.
We note that a unified analysis of the Holstein-Peierls model \cite{Fetherolf_PRX2020,Li_2021} found that, at room temperature and for coupling constants characterizing rubrene, the Holstein term only weakly  affects the dc mobility. Therefore, it is justified for $\mu_{\mathrm{dc}}$ calculation  to neglect the  Holstein term which would otherwise be beyond the applicability of the QC approximation due to the high phonon frequencies. Our findings, hence, indicate that the QC approach provides an optimal method for the calculation of charge mobility which can be incorporated with ab initio electronic structure calculations of OSCs.

%{\it Acknowledgments.}
\section*{Acknowledgments}
We thank V.~Dobrosavljevi\'c for numerous useful discussions.
The authors acknowledge funding
provided by the Institute of Physics Belgrade through
a grant from the Ministry of Science, Technological Development, and Innovation of the Republic of Serbia.

\section*{DATA AVAILABILITY}

The quantum-classical data supporting the findings of this article are openly available in Ref.  \cite{Tanaskovic_Zenodo_QC_multimode}, while the HEOM data were taken directly from Ref. \cite{Jankovic_Zenodo_Peierls}.

%apsrev4-2.bst 2019-01-14 (MD) hand-edited version of apsrev4-1.bst
%Control: key (0)
%Control: author (8) initials jnrlst
%Control: editor formatted (1) identically to author
%Control: production of article title (0) allowed
%Control: page (0) single
%Control: year (1) truncated
%Control: production of eprint (0) enabled
%

%\input{Appendix}
%\bibliographystyle{apsrev4-1}
%\bibliography{refs_multiple_phonons.bib}

\begin{thebibliography}{45}%
\makeatletter
\providecommand \@ifxundefined [1]{%
 \@ifx{#1\undefined}
}%
\providecommand \@ifnum [1]{%
 \ifnum #1\expandafter \@firstoftwo
 \else \expandafter \@secondoftwo
 \fi
}%
\providecommand \@ifx [1]{%
 \ifx #1\expandafter \@firstoftwo
 \else \expandafter \@secondoftwo
 \fi
}%
\providecommand \natexlab [1]{#1}%
\providecommand \enquote  [1]{``#1''}%
\providecommand \bibnamefont  [1]{#1}%
\providecommand \bibfnamefont [1]{#1}%
\providecommand \citenamefont [1]{#1}%
\providecommand \href@noop [0]{\@secondoftwo}%
\providecommand \href [0]{\begingroup \@sanitize@url \@href}%
\providecommand \@href[1]{\@@startlink{#1}\@@href}%
\providecommand \@@href[1]{\endgroup#1\@@endlink}%
\providecommand \@sanitize@url [0]{\catcode `\\12\catcode `\$12\catcode
  `\&12\catcode `\#12\catcode `\^12\catcode `\_12\catcode `\%12\relax}%
\providecommand \@@startlink[1]{}%
\providecommand \@@endlink[0]{}%
\providecommand \url  [0]{\begingroup\@sanitize@url \@url }%
\providecommand \@url [1]{\endgroup\@href {#1}{\urlprefix }}%
\providecommand \urlprefix  [0]{URL }%
\providecommand \Eprint [0]{\href }%
\providecommand \doibase [0]{https://doi.org/}%
\providecommand \selectlanguage [0]{\@gobble}%
\providecommand \bibinfo  [0]{\@secondoftwo}%
\providecommand \bibfield  [0]{\@secondoftwo}%
\providecommand \translation [1]{[#1]}%
\providecommand \BibitemOpen [0]{}%
\providecommand \bibitemStop [0]{}%
\providecommand \bibitemNoStop [0]{.\EOS\space}%
\providecommand \EOS [0]{\spacefactor3000\relax}%
\providecommand \BibitemShut  [1]{\csname bibitem#1\endcsname}%
\let\auto@bib@innerbib\@empty
%</preamble>
\bibitem [{\citenamefont {Coropceanu}\ \emph {et~al.}(2007)\citenamefont
  {Coropceanu}, \citenamefont {Cornil}, \citenamefont {da~Silva~Filho},
  \citenamefont {Olivier}, \citenamefont {Silbey},\ and\ \citenamefont {Br{\'
  e}das}}]{Coropceanu2007}%
  \BibitemOpen
  \bibfield  {author} {\bibinfo {author} {\bibfnamefont {V.}~\bibnamefont
  {Coropceanu}}, \bibinfo {author} {\bibfnamefont {J.}~\bibnamefont {Cornil}},
  \bibinfo {author} {\bibfnamefont {D.~A.}\ \bibnamefont {da~Silva~Filho}},
  \bibinfo {author} {\bibfnamefont {Y.}~\bibnamefont {Olivier}}, \bibinfo
  {author} {\bibfnamefont {R.}~\bibnamefont {Silbey}},\ and\ \bibinfo {author}
  {\bibfnamefont {J.-L.}\ \bibnamefont {Br{\' e}das}},\ }\bibfield  {title}
  {\bibinfo {title} {Charge transport in organic semiconductors},\ }\href
  {https://doi.org/10.1021/cr050140x} {\bibfield  {journal} {\bibinfo
  {journal} {Chem. Rev.}\ }\textbf {\bibinfo {volume} {107}},\ \bibinfo {pages}
  {926} (\bibinfo {year} {2007})}\BibitemShut {NoStop}%
\bibitem [{\citenamefont {Zhugayevych}\ and\ \citenamefont
  {Tretiak}(2015)}]{Zhugayevych2015}%
  \BibitemOpen
  \bibfield  {author} {\bibinfo {author} {\bibfnamefont {A.}~\bibnamefont
  {Zhugayevych}}\ and\ \bibinfo {author} {\bibfnamefont {S.}~\bibnamefont
  {Tretiak}},\ }\bibfield  {title} {\bibinfo {title} {Theoretical description
  of structural and electronic properties of organic photovoltaic materials},\
  }\href
  {https://doi.org/https://doi.org/10.1146/annurev-physchem-040214-121440}
  {\bibfield  {journal} {\bibinfo  {journal} {Annu. Rev. Phys. Chem.}\ }\textbf
  {\bibinfo {volume} {66}},\ \bibinfo {pages} {305} (\bibinfo {year}
  {2015})}\BibitemShut {NoStop}%
\bibitem [{\citenamefont {Fratini}\ \emph {et~al.}(2020)\citenamefont
  {Fratini}, \citenamefont {Nikolka}, \citenamefont {Salleo}, \citenamefont
  {Schweicher},\ and\ \citenamefont {Sirringhaus}}]{Fratini2020}%
  \BibitemOpen
  \bibfield  {author} {\bibinfo {author} {\bibfnamefont {S.}~\bibnamefont
  {Fratini}}, \bibinfo {author} {\bibfnamefont {M.}~\bibnamefont {Nikolka}},
  \bibinfo {author} {\bibfnamefont {A.}~\bibnamefont {Salleo}}, \bibinfo
  {author} {\bibfnamefont {G.}~\bibnamefont {Schweicher}},\ and\ \bibinfo
  {author} {\bibfnamefont {H.}~\bibnamefont {Sirringhaus}},\ }\bibfield
  {title} {\bibinfo {title} {Charge transport in high-mobility conjugated
  polymers and molecular semiconductors},\ }\href
  {https://doi.org/10.1038/s41563-020-0647-2} {\bibfield  {journal} {\bibinfo
  {journal} {Nat. Mater.}\ }\textbf {\bibinfo {volume} {19}},\ \bibinfo {pages}
  {491} (\bibinfo {year} {2020})}\BibitemShut {NoStop}%
\bibitem [{\citenamefont {Nematiaram}\ and\ \citenamefont
  {Troisi}(2020)}]{Nematiaram_2020}%
  \BibitemOpen
  \bibfield  {author} {\bibinfo {author} {\bibfnamefont {T.}~\bibnamefont
  {Nematiaram}}\ and\ \bibinfo {author} {\bibfnamefont {A.}~\bibnamefont
  {Troisi}},\ }\bibfield  {title} {\bibinfo {title} {Modeling charge transport
  in high-mobility molecular semiconductors: Balancing electronic structure and
  quantum dynamics methods with the help of experiments},\ }\href
  {https://doi.org/10.1063/5.0008357} {\bibfield  {journal} {\bibinfo
  {journal} {J. Chem. Phys.}\ }\textbf {\bibinfo {volume} {152}},\ \bibinfo
  {pages} {190902} (\bibinfo {year} {2020})}\BibitemShut {NoStop}%
\bibitem [{\citenamefont {De~Filippis}\ \emph {et~al.}(2015)\citenamefont
  {De~Filippis}, \citenamefont {Cataudella}, \citenamefont {Mishchenko},
  \citenamefont {Nagaosa}, \citenamefont {Fierro},\ and\ \citenamefont
  {de~Candia}}]{2015_DeFilippis}%
  \BibitemOpen
  \bibfield  {author} {\bibinfo {author} {\bibfnamefont {G.}~\bibnamefont
  {De~Filippis}}, \bibinfo {author} {\bibfnamefont {V.}~\bibnamefont
  {Cataudella}}, \bibinfo {author} {\bibfnamefont {A.~S.}\ \bibnamefont
  {Mishchenko}}, \bibinfo {author} {\bibfnamefont {N.}~\bibnamefont {Nagaosa}},
  \bibinfo {author} {\bibfnamefont {A.}~\bibnamefont {Fierro}},\ and\ \bibinfo
  {author} {\bibfnamefont {A.}~\bibnamefont {de~Candia}},\ }\bibfield  {title}
  {\bibinfo {title} {{Crossover from Super- to Subdiffusive Motion and Memory
  Effects in Crystalline Organic Semiconductors}},\ }\href
  {https://doi.org/10.1103/PhysRevLett.114.086601} {\bibfield  {journal}
  {\bibinfo  {journal} {Phys. Rev. Lett.}\ }\textbf {\bibinfo {volume} {114}},\
  \bibinfo {pages} {086601} (\bibinfo {year} {2015})}\BibitemShut {NoStop}%
\bibitem [{\citenamefont {Li}\ \emph {et~al.}(2020)\citenamefont {Li},
  \citenamefont {Ren},\ and\ \citenamefont {Shuai}}]{2020_Li}%
  \BibitemOpen
  \bibfield  {author} {\bibinfo {author} {\bibfnamefont {W.}~\bibnamefont
  {Li}}, \bibinfo {author} {\bibfnamefont {J.}~\bibnamefont {Ren}},\ and\
  \bibinfo {author} {\bibfnamefont {Z.}~\bibnamefont {Shuai}},\ }\bibfield
  {title} {\bibinfo {title} {{Finite-temperature TD-DMRG for the carrier
  mobility of organic semiconductors}},\ }\href
  {https://doi.org/10.1021/acs.jpclett.0c01072} {\bibfield  {journal} {\bibinfo
   {journal} {J. Phys. Chem. Lett.}\ }\textbf {\bibinfo {volume} {11}},\
  \bibinfo {pages} {4930} (\bibinfo {year} {2020})}\BibitemShut {NoStop}%
\bibitem [{\citenamefont {Bertini}\ \emph {et~al.}(2021)\citenamefont
  {Bertini}, \citenamefont {Heidrich-Meisner}, \citenamefont {Karrasch},
  \citenamefont {Prosen}, \citenamefont {Steinigeweg},\ and\ \citenamefont
  {\ifmmode \check{Z}\else \v{Z}\fi{}nidari\ifmmode~\check{c}\else
  \v{c}\fi{}}}]{2021_Bertini}%
  \BibitemOpen
  \bibfield  {author} {\bibinfo {author} {\bibfnamefont {B.}~\bibnamefont
  {Bertini}}, \bibinfo {author} {\bibfnamefont {F.}~\bibnamefont
  {Heidrich-Meisner}}, \bibinfo {author} {\bibfnamefont {C.}~\bibnamefont
  {Karrasch}}, \bibinfo {author} {\bibfnamefont {T.}~\bibnamefont {Prosen}},
  \bibinfo {author} {\bibfnamefont {R.}~\bibnamefont {Steinigeweg}},\ and\
  \bibinfo {author} {\bibfnamefont {M.}~\bibnamefont {\ifmmode \check{Z}\else
  \v{Z}\fi{}nidari\ifmmode~\check{c}\else \v{c}\fi{}}},\ }\bibfield  {title}
  {\bibinfo {title} {Finite-temperature transport in one-dimensional quantum
  lattice models},\ }\href {https://doi.org/10.1103/RevModPhys.93.025003}
  {\bibfield  {journal} {\bibinfo  {journal} {Rev. Mod. Phys.}\ }\textbf
  {\bibinfo {volume} {93}},\ \bibinfo {pages} {025003} (\bibinfo {year}
  {2021})}\BibitemShut {NoStop}%
\bibitem [{\citenamefont {Ostmeyer}\ \emph {et~al.}(2024)\citenamefont
  {Ostmeyer}, \citenamefont {Nematiaram}, \citenamefont {Troisi},\ and\
  \citenamefont {Buividovich}}]{2024_Ostmeyer}%
  \BibitemOpen
  \bibfield  {author} {\bibinfo {author} {\bibfnamefont {J.}~\bibnamefont
  {Ostmeyer}}, \bibinfo {author} {\bibfnamefont {T.}~\bibnamefont
  {Nematiaram}}, \bibinfo {author} {\bibfnamefont {A.}~\bibnamefont {Troisi}},\
  and\ \bibinfo {author} {\bibfnamefont {P.}~\bibnamefont {Buividovich}},\
  }\bibfield  {title} {\bibinfo {title} {First-principles quantum monte carlo
  study of charge-carrier mobility in organic molecular semiconductors},\
  }\href {https://doi.org/10.1103/PhysRevApplied.22.L031004} {\bibfield
  {journal} {\bibinfo  {journal} {Phys. Rev. Appl.}\ }\textbf {\bibinfo
  {volume} {22}},\ \bibinfo {pages} {L031004} (\bibinfo {year}
  {2024})}\BibitemShut {NoStop}%
\bibitem [{\citenamefont {Buividovich}\ \emph {et~al.}(2024)\citenamefont
  {Buividovich}, \citenamefont {Ostmeyer},\ and\ \citenamefont
  {Troisi}}]{Buividovich:2024MY}%
  \BibitemOpen
  \bibfield  {author} {\bibinfo {author} {\bibfnamefont {P.}~\bibnamefont
  {Buividovich}}, \bibinfo {author} {\bibfnamefont {J.}~\bibnamefont
  {Ostmeyer}},\ and\ \bibinfo {author} {\bibfnamefont {A.}~\bibnamefont
  {Troisi}},\ }\bibfield  {title} {\bibinfo {title} {{High-precision Quantum
  Monte-Carlo study of charge transport in a lattice model of molecular organic
  semiconductors}},\ }\href {https://doi.org/10.22323/1.466.0067} {\bibfield
  {journal} {\bibinfo  {journal} {PoS}\ }\textbf {\bibinfo {volume}
  {LATTICE2024}},\ \bibinfo {pages} {067} (\bibinfo {year} {2024})}\BibitemShut
  {NoStop}%
\bibitem [{\citenamefont {Fratini}\ \emph {et~al.}(2016)\citenamefont
  {Fratini}, \citenamefont {Mayou},\ and\ \citenamefont
  {Ciuchi}}]{Fratini2016}%
  \BibitemOpen
  \bibfield  {author} {\bibinfo {author} {\bibfnamefont {S.}~\bibnamefont
  {Fratini}}, \bibinfo {author} {\bibfnamefont {D.}~\bibnamefont {Mayou}},\
  and\ \bibinfo {author} {\bibfnamefont {S.}~\bibnamefont {Ciuchi}},\
  }\bibfield  {title} {\bibinfo {title} {The transient localization scenario
  for charge transport in crystalline organic materials},\ }\href
  {https://doi.org/https://doi.org/10.1002/adfm.201502386} {\bibfield
  {journal} {\bibinfo  {journal} {Adv. Funct. Mater.}\ }\textbf {\bibinfo
  {volume} {26}},\ \bibinfo {pages} {2292} (\bibinfo {year}
  {2016})}\BibitemShut {NoStop}%
\bibitem [{\citenamefont {Ciuchi}\ \emph {et~al.}(2011)\citenamefont {Ciuchi},
  \citenamefont {Fratini},\ and\ \citenamefont {Mayou}}]{Ciuchi_PRB2011}%
  \BibitemOpen
  \bibfield  {author} {\bibinfo {author} {\bibfnamefont {S.}~\bibnamefont
  {Ciuchi}}, \bibinfo {author} {\bibfnamefont {S.}~\bibnamefont {Fratini}},\
  and\ \bibinfo {author} {\bibfnamefont {D.}~\bibnamefont {Mayou}},\ }\bibfield
   {title} {\bibinfo {title} {Transient localization in crystalline organic
  semiconductors},\ }\href {https://doi.org/10.1103/PhysRevB.83.081202}
  {\bibfield  {journal} {\bibinfo  {journal} {Phys. Rev. B}\ }\textbf {\bibinfo
  {volume} {83}},\ \bibinfo {pages} {081202} (\bibinfo {year}
  {2011})}\BibitemShut {NoStop}%
\bibitem [{\citenamefont {Fratini}\ \emph {et~al.}(2017)\citenamefont
  {Fratini}, \citenamefont {Ciuchi}, \citenamefont {Mayou}, \citenamefont
  {de~Laissardi{\` e}re},\ and\ \citenamefont {Troisi}}]{Fratini_NatMat2017}%
  \BibitemOpen
  \bibfield  {author} {\bibinfo {author} {\bibfnamefont {S.}~\bibnamefont
  {Fratini}}, \bibinfo {author} {\bibfnamefont {S.}~\bibnamefont {Ciuchi}},
  \bibinfo {author} {\bibfnamefont {D.}~\bibnamefont {Mayou}}, \bibinfo
  {author} {\bibfnamefont {G.}~\bibnamefont {de~Laissardi{\` e}re}},\ and\
  \bibinfo {author} {\bibfnamefont {A.}~\bibnamefont {Troisi}},\ }\bibfield
  {title} {\bibinfo {title} {A map of high-mobility molecular semiconductors},\
  }\href {https://doi.org/10.1038/nmat4970} {\bibfield  {journal} {\bibinfo
  {journal} {Nat. Mater.}\ }\textbf {\bibinfo {volume} {16}},\ \bibinfo {pages}
  {998} (\bibinfo {year} {2017})}\BibitemShut {NoStop}%
\bibitem [{\citenamefont {Troisi}\ and\ \citenamefont
  {Orlandi}(2006)}]{Troisi_PRL2006}%
  \BibitemOpen
  \bibfield  {author} {\bibinfo {author} {\bibfnamefont {A.}~\bibnamefont
  {Troisi}}\ and\ \bibinfo {author} {\bibfnamefont {G.}~\bibnamefont
  {Orlandi}},\ }\bibfield  {title} {\bibinfo {title} {{Charge-Transport Regime
  of Crystalline Organic Semiconductors: Diffusion Limited by Thermal
  Off-Diagonal Electronic Disorder}},\ }\href
  {https://doi.org/10.1103/PhysRevLett.96.086601} {\bibfield  {journal}
  {\bibinfo  {journal} {Phys. Rev. Lett.}\ }\textbf {\bibinfo {volume} {96}},\
  \bibinfo {pages} {086601} (\bibinfo {year} {2006})}\BibitemShut {NoStop}%
\bibitem [{\citenamefont {Parandekar}\ and\ \citenamefont
  {Tully}(2005)}]{Parandekar_2005}%
  \BibitemOpen
  \bibfield  {author} {\bibinfo {author} {\bibfnamefont {P.~V.}\ \bibnamefont
  {Parandekar}}\ and\ \bibinfo {author} {\bibfnamefont {J.~C.}\ \bibnamefont
  {Tully}},\ }\bibfield  {title} {\bibinfo {title} {Mixed quantum-classical
  equilibrium},\ }\href {https://doi.org/10.1063/1.1856460} {\bibfield
  {journal} {\bibinfo  {journal} {J. Chem. Phys.}\ }\textbf {\bibinfo {volume}
  {122}},\ \bibinfo {pages} {094102} (\bibinfo {year} {2005})}\BibitemShut
  {NoStop}%
\bibitem [{\citenamefont {Tully}(2023)}]{Tully_2023}%
  \BibitemOpen
  \bibfield  {author} {\bibinfo {author} {\bibfnamefont {J.~C.}\ \bibnamefont
  {Tully}},\ }\bibfield  {title} {\bibinfo {title} {Ehrenfest dynamics with
  quantum mechanical nuclei},\ }\href
  {https://doi.org/https://doi.org/10.1016/j.cplett.2023.140396} {\bibfield
  {journal} {\bibinfo  {journal} {Chem. Phys. Lett.}\ }\textbf {\bibinfo
  {volume} {816}},\ \bibinfo {pages} {140396} (\bibinfo {year}
  {2023})}\BibitemShut {NoStop}%
\bibitem [{\citenamefont {Fetherolf}\ \emph {et~al.}(2020)\citenamefont
  {Fetherolf}, \citenamefont {Gole\ifmmode~\check{z}\else \v{z}\fi{}},\ and\
  \citenamefont {Berkelbach}}]{Fetherolf_PRX2020}%
  \BibitemOpen
  \bibfield  {author} {\bibinfo {author} {\bibfnamefont {J.~H.}\ \bibnamefont
  {Fetherolf}}, \bibinfo {author} {\bibfnamefont {D.}~\bibnamefont
  {Gole\ifmmode~\check{z}\else \v{z}\fi{}}},\ and\ \bibinfo {author}
  {\bibfnamefont {T.~C.}\ \bibnamefont {Berkelbach}},\ }\bibfield  {title}
  {\bibinfo {title} {A unification of the {H}olstein polaron and dynamic
  disorder pictures of charge transport in organic crystals},\ }\href
  {https://doi.org/10.1103/PhysRevX.10.021062} {\bibfield  {journal} {\bibinfo
  {journal} {Phys. Rev. X}\ }\textbf {\bibinfo {volume} {10}},\ \bibinfo
  {pages} {021062} (\bibinfo {year} {2020})}\BibitemShut {NoStop}%
\bibitem [{\citenamefont {Runeson}\ \emph {et~al.}(2024)\citenamefont
  {Runeson}, \citenamefont {Drayton},\ and\ \citenamefont
  {Manolopoulos}}]{Runeson_2024}%
  \BibitemOpen
  \bibfield  {author} {\bibinfo {author} {\bibfnamefont {J.~E.}\ \bibnamefont
  {Runeson}}, \bibinfo {author} {\bibfnamefont {T.~J.~G.}\ \bibnamefont
  {Drayton}},\ and\ \bibinfo {author} {\bibfnamefont {D.~E.}\ \bibnamefont
  {Manolopoulos}},\ }\bibfield  {title} {\bibinfo {title} {Charge transport in
  organic semiconductors from the mapping approach to surface hopping},\ }\href
  {https://doi.org/10.1063/5.0226001} {\bibfield  {journal} {\bibinfo
  {journal} {J. Chem. Phys.}\ }\textbf {\bibinfo {volume} {161}},\ \bibinfo
  {pages} {144102} (\bibinfo {year} {2024})}\BibitemShut {NoStop}%
\bibitem [{\citenamefont {Mitri\ifmmode~\acute{c}\else \'{c}\fi{}}\ \emph
  {et~al.}(2025)\citenamefont {Mitri\ifmmode~\acute{c}\else \'{c}\fi{}},
  \citenamefont {Dobrosavljevi\ifmmode~\acute{c}\else \'{c}\fi{}},\ and\
  \citenamefont {Tanaskovi\ifmmode~\acute{c}\else
  \'{c}\fi{}}}]{Mitric_Precursors_2025}%
  \BibitemOpen
  \bibfield  {author} {\bibinfo {author} {\bibfnamefont {P.}~\bibnamefont
  {Mitri\ifmmode~\acute{c}\else \'{c}\fi{}}}, \bibinfo {author} {\bibfnamefont
  {V.}~\bibnamefont {Dobrosavljevi\ifmmode~\acute{c}\else \'{c}\fi{}}},\ and\
  \bibinfo {author} {\bibfnamefont {D.}~\bibnamefont
  {Tanaskovi\ifmmode~\acute{c}\else \'{c}\fi{}}},\ }\bibfield  {title}
  {\bibinfo {title} {{Precursors to Anderson localization in the Holstein
  model: Quantum and quantum-classical solutions}},\ }\href
  {https://doi.org/10.1103/PhysRevB.111.L161105} {\bibfield  {journal}
  {\bibinfo  {journal} {Phys. Rev. B}\ }\textbf {\bibinfo {volume} {111}},\
  \bibinfo {pages} {L161105} (\bibinfo {year} {2025})}\BibitemShut {NoStop}%
\bibitem [{\citenamefont {Jankovi\ifmmode~\acute{c}\else
  \'{c}\fi{}}(2025{\natexlab{a}})}]{Jankovic_DEOM_2025}%
  \BibitemOpen
  \bibfield  {author} {\bibinfo {author} {\bibfnamefont {V.}~\bibnamefont
  {Jankovi\ifmmode~\acute{c}\else \'{c}\fi{}}},\ }\bibfield  {title} {\bibinfo
  {title} {Numerically "exact" charge transport dynamics in a dissipative
  electron-phonon model rationalizing the success of the transient localization
  scenario},\ }\href {https://doi.org/10.1063/5.0299359} {\bibfield  {journal}
  {\bibinfo  {journal} {J. Chem. Phys.}\ }\textbf {\bibinfo {volume} {163}},\
  \bibinfo {pages} {194116} (\bibinfo {year} {2025}{\natexlab{a}})}\BibitemShut
  {NoStop}%
\bibitem [{\citenamefont {Peluzo}\ \emph {et~al.}(2025)\citenamefont {Peluzo},
  \citenamefont {Meena}, \citenamefont {Catalano}, \citenamefont {Schweicher},\
  and\ \citenamefont {Ruggiero}}]{Peluzo_2025}%
  \BibitemOpen
  \bibfield  {author} {\bibinfo {author} {\bibfnamefont {B.~M. T.~C.}\
  \bibnamefont {Peluzo}}, \bibinfo {author} {\bibfnamefont {R.}~\bibnamefont
  {Meena}}, \bibinfo {author} {\bibfnamefont {L.}~\bibnamefont {Catalano}},
  \bibinfo {author} {\bibfnamefont {G.}~\bibnamefont {Schweicher}},\ and\
  \bibinfo {author} {\bibfnamefont {M.~T.}\ \bibnamefont {Ruggiero}},\
  }\bibfield  {title} {\bibinfo {title} {Exploring the interplay of lattice
  dynamics and charge transport in organic semiconductors: Progress toward
  rational phonon engineering},\ }\href
  {https://doi.org/https://doi.org/10.1002/anie.202507566} {\bibfield
  {journal} {\bibinfo  {journal} {Angew. Chem. Int. Ed.}\ }\textbf {\bibinfo
  {volume} {64}},\ \bibinfo {pages} {e202507566} (\bibinfo {year}
  {2025})}\BibitemShut {NoStop}%
\bibitem [{\citenamefont {Schweicher}\ \emph {et~al.}(2019)\citenamefont
  {Schweicher}, \citenamefont {D'Avino}, \citenamefont {Ruggiero},
  \citenamefont {Harkin}, \citenamefont {Broch}, \citenamefont
  {Venkateshvaran}, \citenamefont {Liu}, \citenamefont {Richard}, \citenamefont
  {Ruzi{\' e}}, \citenamefont {Armstrong}, \citenamefont {Kennedy},
  \citenamefont {Shankland}, \citenamefont {Takimiya}, \citenamefont {Geerts},
  \citenamefont {Zeitler}, \citenamefont {Fratini},\ and\ \citenamefont
  {Sirringhaus}}]{Schweicher_2019}%
  \BibitemOpen
  \bibfield  {author} {\bibinfo {author} {\bibfnamefont {G.}~\bibnamefont
  {Schweicher}}, \bibinfo {author} {\bibfnamefont {G.}~\bibnamefont {D'Avino}},
  \bibinfo {author} {\bibfnamefont {M.~T.}\ \bibnamefont {Ruggiero}}, \bibinfo
  {author} {\bibfnamefont {D.~J.}\ \bibnamefont {Harkin}}, \bibinfo {author}
  {\bibfnamefont {K.}~\bibnamefont {Broch}}, \bibinfo {author} {\bibfnamefont
  {D.}~\bibnamefont {Venkateshvaran}}, \bibinfo {author} {\bibfnamefont
  {G.}~\bibnamefont {Liu}}, \bibinfo {author} {\bibfnamefont {A.}~\bibnamefont
  {Richard}}, \bibinfo {author} {\bibfnamefont {C.}~\bibnamefont {Ruzi{\' e}}},
  \bibinfo {author} {\bibfnamefont {J.}~\bibnamefont {Armstrong}}, \bibinfo
  {author} {\bibfnamefont {A.~R.}\ \bibnamefont {Kennedy}}, \bibinfo {author}
  {\bibfnamefont {K.}~\bibnamefont {Shankland}}, \bibinfo {author}
  {\bibfnamefont {K.}~\bibnamefont {Takimiya}}, \bibinfo {author}
  {\bibfnamefont {Y.~H.}\ \bibnamefont {Geerts}}, \bibinfo {author}
  {\bibfnamefont {J.~A.}\ \bibnamefont {Zeitler}}, \bibinfo {author}
  {\bibfnamefont {S.}~\bibnamefont {Fratini}},\ and\ \bibinfo {author}
  {\bibfnamefont {H.}~\bibnamefont {Sirringhaus}},\ }\bibfield  {title}
  {\bibinfo {title} {Chasing the “killer” phonon mode for the rational
  design of low-disorder, high-mobility molecular semiconductors},\ }\href
  {https://doi.org/https://doi.org/10.1002/adma.201902407} {\bibfield
  {journal} {\bibinfo  {journal} {Adv. Mater.}\ }\textbf {\bibinfo {volume}
  {31}},\ \bibinfo {pages} {1902407} (\bibinfo {year} {2019})}\BibitemShut
  {NoStop}%
\bibitem [{\citenamefont {Jankovi\ifmmode~\acute{c}\else
  \'{c}\fi{}}(2025{\natexlab{b}})}]{Jankovic_HEOMII_2025}%
  \BibitemOpen
  \bibfield  {author} {\bibinfo {author} {\bibfnamefont {V.}~\bibnamefont
  {Jankovi\ifmmode~\acute{c}\else \'{c}\fi{}}},\ }\bibfield  {title} {\bibinfo
  {title} {Charge transport limited by nonlocal electron-phonon interaction.
  {II}. {N}umerically exact quantum dynamics in the slow-phonon regime},\
  }\href {https://doi.org/10.1103/f56z-h612} {\bibfield  {journal} {\bibinfo
  {journal} {Phys. Rev. B}\ }\textbf {\bibinfo {volume} {112}},\ \bibinfo
  {pages} {035112} (\bibinfo {year} {2025}{\natexlab{b}})}\BibitemShut
  {NoStop}%
\bibitem [{\citenamefont {Mitri\ifmmode~\acute{c}\else
  \'{c}\fi{}}(2025)}]{Mitric_QT_2025}%
  \BibitemOpen
  \bibfield  {author} {\bibinfo {author} {\bibfnamefont {P.}~\bibnamefont
  {Mitri\ifmmode~\acute{c}\else \'{c}\fi{}}},\ }\bibfield  {title} {\bibinfo
  {title} {{Dynamical quantum typicality: Simple method for investigating
  transport properties applied to the Holstein model}},\ }\href
  {https://doi.org/10.1103/PhysRevB.111.195140} {\bibfield  {journal} {\bibinfo
   {journal} {Phys. Rev. B}\ }\textbf {\bibinfo {volume} {111}},\ \bibinfo
  {pages} {195140} (\bibinfo {year} {2025})}\BibitemShut {NoStop}%
\bibitem [{\citenamefont {Jankovi\ifmmode~\acute{c}\else \'{c}\fi{}}\ \emph
  {et~al.}(2024)\citenamefont {Jankovi\ifmmode~\acute{c}\else \'{c}\fi{}},
  \citenamefont {Mitri\ifmmode~\acute{c}\else \'{c}\fi{}}, \citenamefont
  {Tanaskovi\ifmmode~\acute{c}\else \'{c}\fi{}},\ and\ \citenamefont
  {Vukmirovi\ifmmode~\acute{c}\else \'{c}\fi{}}}]{Jankovic_2024}%
  \BibitemOpen
  \bibfield  {author} {\bibinfo {author} {\bibfnamefont {V.}~\bibnamefont
  {Jankovi\ifmmode~\acute{c}\else \'{c}\fi{}}}, \bibinfo {author}
  {\bibfnamefont {P.}~\bibnamefont {Mitri\ifmmode~\acute{c}\else \'{c}\fi{}}},
  \bibinfo {author} {\bibfnamefont {D.}~\bibnamefont
  {Tanaskovi\ifmmode~\acute{c}\else \'{c}\fi{}}},\ and\ \bibinfo {author}
  {\bibfnamefont {N.}~\bibnamefont {Vukmirovi\ifmmode~\acute{c}\else
  \'{c}\fi{}}},\ }\bibfield  {title} {\bibinfo {title} {Vertex corrections to
  conductivity in the {H}olstein model: A numerical-analytical study},\ }\href
  {https://doi.org/10.1103/PhysRevB.109.214312} {\bibfield  {journal} {\bibinfo
   {journal} {Phys. Rev. B}\ }\textbf {\bibinfo {volume} {109}},\ \bibinfo
  {pages} {214312} (\bibinfo {year} {2024})}\BibitemShut {NoStop}%
\bibitem [{\citenamefont {Jankovi\ifmmode~\acute{c}\else
  \'{c}\fi{}}(2023)}]{Jankovic_2023}%
  \BibitemOpen
  \bibfield  {author} {\bibinfo {author} {\bibfnamefont {V.}~\bibnamefont
  {Jankovi\ifmmode~\acute{c}\else \'{c}\fi{}}},\ }\bibfield  {title} {\bibinfo
  {title} {Holstein polaron transport from numerically "exact" real-time
  quantum dynamics simulations},\ }\href {https://doi.org/10.1063/5.0165532}
  {\bibfield  {journal} {\bibinfo  {journal} {J. Chem. Phys.}\ }\textbf
  {\bibinfo {volume} {159}},\ \bibinfo {pages} {094113} (\bibinfo {year}
  {2023})}\BibitemShut {NoStop}%
\bibitem [{\citenamefont {Mahan}(1990)}]{Mahan}%
  \BibitemOpen
  \bibfield  {author} {\bibinfo {author} {\bibfnamefont {G.}~\bibnamefont
  {Mahan}},\ }\href@noop {} {\emph {\bibinfo {title} {Many-Particle Physics}}}\
  (\bibinfo  {publisher} {Plenum, New York},\ \bibinfo {year}
  {1990})\BibitemShut {NoStop}%
\bibitem [{\citenamefont {Kubo}(1957)}]{Kubo_1957}%
  \BibitemOpen
  \bibfield  {author} {\bibinfo {author} {\bibfnamefont {R.}~\bibnamefont
  {Kubo}},\ }\bibfield  {title} {\bibinfo {title} {Statistical-mechanical
  theory of irreversible processes. {I}. {G}eneral theory and simple
  applications to magnetic and conduction problems},\ }\href
  {https://doi.org/10.1143/JPSJ.12.570} {\bibfield  {journal} {\bibinfo
  {journal} {J. Phys. Soc. Jpn.}\ }\textbf {\bibinfo {volume} {12}},\ \bibinfo
  {pages} {570} (\bibinfo {year} {1957})}\BibitemShut {NoStop}%
\bibitem [{\citenamefont {Fratini}\ and\ \citenamefont
  {Ciuchi}(2020)}]{Fratini_PRR2020}%
  \BibitemOpen
  \bibfield  {author} {\bibinfo {author} {\bibfnamefont {S.}~\bibnamefont
  {Fratini}}\ and\ \bibinfo {author} {\bibfnamefont {S.}~\bibnamefont
  {Ciuchi}},\ }\bibfield  {title} {\bibinfo {title} {Dynamical localization
  corrections to band transport},\ }\href
  {https://doi.org/10.1103/PhysRevResearch.2.013001} {\bibfield  {journal}
  {\bibinfo  {journal} {Phys. Rev. Res.}\ }\textbf {\bibinfo {volume} {2}},\
  \bibinfo {pages} {013001} (\bibinfo {year} {2020})}\BibitemShut {NoStop}%
\bibitem [{\citenamefont {Troisi}(2007)}]{Troisi_2007}%
  \BibitemOpen
  \bibfield  {author} {\bibinfo {author} {\bibfnamefont {A.}~\bibnamefont
  {Troisi}},\ }\bibfield  {title} {\bibinfo {title} {Prediction of the absolute
  charge mobility of molecular semiconductors: the case of rubrene},\ }\href
  {https://doi.org/https://doi.org/10.1002/adma.200700550} {\bibfield
  {journal} {\bibinfo  {journal} {Advanced Materials}\ }\textbf {\bibinfo
  {volume} {19}},\ \bibinfo {pages} {2000} (\bibinfo {year}
  {2007})}\BibitemShut {NoStop}%
\bibitem [{\citenamefont {Girlando}\ \emph {et~al.}(2010)\citenamefont
  {Girlando}, \citenamefont {Grisanti}, \citenamefont {Masino}, \citenamefont
  {Bilotti}, \citenamefont {Brillante}, \citenamefont {Della~Valle},\ and\
  \citenamefont {Venuti}}]{Girlando_2010}%
  \BibitemOpen
  \bibfield  {author} {\bibinfo {author} {\bibfnamefont {A.}~\bibnamefont
  {Girlando}}, \bibinfo {author} {\bibfnamefont {L.}~\bibnamefont {Grisanti}},
  \bibinfo {author} {\bibfnamefont {M.}~\bibnamefont {Masino}}, \bibinfo
  {author} {\bibfnamefont {I.}~\bibnamefont {Bilotti}}, \bibinfo {author}
  {\bibfnamefont {A.}~\bibnamefont {Brillante}}, \bibinfo {author}
  {\bibfnamefont {R.~G.}\ \bibnamefont {Della~Valle}},\ and\ \bibinfo {author}
  {\bibfnamefont {E.}~\bibnamefont {Venuti}},\ }\bibfield  {title} {\bibinfo
  {title} {Peierls and holstein carrier-phonon coupling in crystalline
  rubrene},\ }\href {https://doi.org/10.1103/PhysRevB.82.035208} {\bibfield
  {journal} {\bibinfo  {journal} {Phys. Rev. B}\ }\textbf {\bibinfo {volume}
  {82}},\ \bibinfo {pages} {035208} (\bibinfo {year} {2010})}\BibitemShut
  {NoStop}%
\bibitem [{\citenamefont {Ordej\'on}\ \emph {et~al.}(2017)\citenamefont
  {Ordej\'on}, \citenamefont {Boskovic}, \citenamefont {Panhans},\ and\
  \citenamefont {Ortmann}}]{Ordejon_2017}%
  \BibitemOpen
  \bibfield  {author} {\bibinfo {author} {\bibfnamefont {P.}~\bibnamefont
  {Ordej\'on}}, \bibinfo {author} {\bibfnamefont {D.}~\bibnamefont {Boskovic}},
  \bibinfo {author} {\bibfnamefont {M.}~\bibnamefont {Panhans}},\ and\ \bibinfo
  {author} {\bibfnamefont {F.}~\bibnamefont {Ortmann}},\ }\bibfield  {title}
  {\bibinfo {title} {Ab initio study of electron-phonon coupling in rubrene},\
  }\href {https://doi.org/10.1103/PhysRevB.96.035202} {\bibfield  {journal}
  {\bibinfo  {journal} {Phys. Rev. B}\ }\textbf {\bibinfo {volume} {96}},\
  \bibinfo {pages} {035202} (\bibinfo {year} {2017})}\BibitemShut {NoStop}%
\bibitem [{\citenamefont {Jankovi\ifmmode~\acute{c}\else
  \'{c}\fi{}}()}]{Jankovic_Zenodo_Peierls}%
  \BibitemOpen
  \bibfield  {author} {\bibinfo {author} {\bibfnamefont {V.}~\bibnamefont
  {Jankovi\ifmmode~\acute{c}\else \'{c}\fi{}}},\ }\href
  {https://doi.org/10.5281/zenodo.14637019} {\bibinfo {title} {Numerical
  investigation of transport properties of the one-dimensional {P}eierls model
  based on the hierarchical equations of motion [{D}ata set]}},\ \bibinfo
  {note} {{Z}enodo (2025)}\BibitemShut {NoStop}%
\bibitem [{\citenamefont {Fischer}\ \emph {et~al.}(2006)\citenamefont
  {Fischer}, \citenamefont {Dressel}, \citenamefont {Gompf}, \citenamefont
  {Tripathi},\ and\ \citenamefont {Pflaum}}]{Fischer_2006}%
  \BibitemOpen
  \bibfield  {author} {\bibinfo {author} {\bibfnamefont {M.}~\bibnamefont
  {Fischer}}, \bibinfo {author} {\bibfnamefont {M.}~\bibnamefont {Dressel}},
  \bibinfo {author} {\bibfnamefont {B.}~\bibnamefont {Gompf}}, \bibinfo
  {author} {\bibfnamefont {A.~K.}\ \bibnamefont {Tripathi}},\ and\ \bibinfo
  {author} {\bibfnamefont {J.}~\bibnamefont {Pflaum}},\ }\bibfield  {title}
  {\bibinfo {title} {Infrared spectroscopy on the charge accumulation layer in
  rubrene single crystals},\ }\href {https://doi.org/10.1063/1.2370743}
  {\bibfield  {journal} {\bibinfo  {journal} {Appl. Phys. Lett.}\ }\textbf
  {\bibinfo {volume} {89}},\ \bibinfo {pages} {182103} (\bibinfo {year}
  {2006})}\BibitemShut {NoStop}%
\bibitem [{\citenamefont {Li}\ \emph {et~al.}(2007)\citenamefont {Li},
  \citenamefont {Podzorov}, \citenamefont {Sai}, \citenamefont {Martin},
  \citenamefont {Gershenson}, \citenamefont {Di~Ventra},\ and\ \citenamefont
  {Basov}}]{Li_2007}%
  \BibitemOpen
  \bibfield  {author} {\bibinfo {author} {\bibfnamefont {Z.~Q.}\ \bibnamefont
  {Li}}, \bibinfo {author} {\bibfnamefont {V.}~\bibnamefont {Podzorov}},
  \bibinfo {author} {\bibfnamefont {N.}~\bibnamefont {Sai}}, \bibinfo {author}
  {\bibfnamefont {M.~C.}\ \bibnamefont {Martin}}, \bibinfo {author}
  {\bibfnamefont {M.~E.}\ \bibnamefont {Gershenson}}, \bibinfo {author}
  {\bibfnamefont {M.}~\bibnamefont {Di~Ventra}},\ and\ \bibinfo {author}
  {\bibfnamefont {D.~N.}\ \bibnamefont {Basov}},\ }\bibfield  {title} {\bibinfo
  {title} {{Light Quasiparticles Dominate Electronic Transport in Molecular
  Crystal Field-Effect Transistors}},\ }\href
  {https://doi.org/10.1103/PhysRevLett.99.016403} {\bibfield  {journal}
  {\bibinfo  {journal} {Phys. Rev. Lett.}\ }\textbf {\bibinfo {volume} {99}},\
  \bibinfo {pages} {016403} (\bibinfo {year} {2007})}\BibitemShut {NoStop}%
\bibitem [{\citenamefont {Fetherolf}\ \emph {et~al.}(2023)\citenamefont
  {Fetherolf}, \citenamefont {Shih},\ and\ \citenamefont
  {Berkelbach}}]{Fetherolf_2023}%
  \BibitemOpen
  \bibfield  {author} {\bibinfo {author} {\bibfnamefont {J.~H.}\ \bibnamefont
  {Fetherolf}}, \bibinfo {author} {\bibfnamefont {P.}~\bibnamefont {Shih}},\
  and\ \bibinfo {author} {\bibfnamefont {T.~C.}\ \bibnamefont {Berkelbach}},\
  }\bibfield  {title} {\bibinfo {title} {Conductivity of an electron coupled to
  anharmonic phonons: Quantum-classical simulations and comparison of
  approximations},\ }\href {https://doi.org/10.1103/PhysRevB.107.064304}
  {\bibfield  {journal} {\bibinfo  {journal} {Phys. Rev. B}\ }\textbf {\bibinfo
  {volume} {107}},\ \bibinfo {pages} {064304} (\bibinfo {year}
  {2023})}\BibitemShut {NoStop}%
\bibitem [{\citenamefont {Wang}\ \emph {et~al.}(2011)\citenamefont {Wang},
  \citenamefont {Beljonne}, \citenamefont {Chen},\ and\ \citenamefont
  {Shi}}]{Wang_2011}%
  \BibitemOpen
  \bibfield  {author} {\bibinfo {author} {\bibfnamefont {L.}~\bibnamefont
  {Wang}}, \bibinfo {author} {\bibfnamefont {D.}~\bibnamefont {Beljonne}},
  \bibinfo {author} {\bibfnamefont {L.}~\bibnamefont {Chen}},\ and\ \bibinfo
  {author} {\bibfnamefont {Q.}~\bibnamefont {Shi}},\ }\bibfield  {title}
  {\bibinfo {title} {Mixed quantum-classical simulations of charge transport in
  organic materials: Numerical benchmark of the {S}u-{S}chrieffer-{H}eeger
  model},\ }\href {https://doi.org/10.1063/1.3604561} {\bibfield  {journal}
  {\bibinfo  {journal} {J. Chem. Phys.}\ }\textbf {\bibinfo {volume} {134}},\
  \bibinfo {pages} {244116} (\bibinfo {year} {2011})}\BibitemShut {NoStop}%
\bibitem [{\citenamefont {ten Brink}\ \emph {et~al.}(2022)\citenamefont {ten
  Brink}, \citenamefont {Gräber}, \citenamefont {Hopjan}, \citenamefont
  {Jansen}, \citenamefont {Stolpp}, \citenamefont {Heidrich-Meisner},\ and\
  \citenamefont {Blöchl}}]{Brink_2022}%
  \BibitemOpen
  \bibfield  {author} {\bibinfo {author} {\bibfnamefont {M.}~\bibnamefont {ten
  Brink}}, \bibinfo {author} {\bibfnamefont {S.}~\bibnamefont {Gräber}},
  \bibinfo {author} {\bibfnamefont {M.}~\bibnamefont {Hopjan}}, \bibinfo
  {author} {\bibfnamefont {D.}~\bibnamefont {Jansen}}, \bibinfo {author}
  {\bibfnamefont {J.}~\bibnamefont {Stolpp}}, \bibinfo {author} {\bibfnamefont
  {F.}~\bibnamefont {Heidrich-Meisner}},\ and\ \bibinfo {author} {\bibfnamefont
  {P.~E.}\ \bibnamefont {Blöchl}},\ }\bibfield  {title} {\bibinfo {title}
  {Real-time non-adiabatic dynamics in the one-dimensional {H}olstein model:
  Trajectory-based vs exact methods},\ }\href
  {https://doi.org/10.1063/5.0092063} {\bibfield  {journal} {\bibinfo
  {journal} {The Journal of Chemical Physics}\ }\textbf {\bibinfo {volume}
  {156}},\ \bibinfo {pages} {234109} (\bibinfo {year} {2022})}\BibitemShut
  {NoStop}%
\bibitem [{\citenamefont {Menzler}\ \emph {et~al.}(2025)\citenamefont
  {Menzler}, \citenamefont {Mondal},\ and\ \citenamefont
  {Heidrich-Meisner}}]{Menzler_2025}%
  \BibitemOpen
  \bibfield  {author} {\bibinfo {author} {\bibfnamefont {H.~G.}\ \bibnamefont
  {Menzler}}, \bibinfo {author} {\bibfnamefont {S.}~\bibnamefont {Mondal}},\
  and\ \bibinfo {author} {\bibfnamefont {F.}~\bibnamefont {Heidrich-Meisner}},\
  }\href {https://arxiv.org/abs/2512.10899} {\bibinfo {title} {Hybrid
  quantum-classical matrix-product state and {L}anczos methods for
  electron-phonon systems with strong electronic correlations: {A}pplication to
  disordered systems coupled to {E}instein phonons}} (\bibinfo {year} {2025}),\
  \Eprint {https://arxiv.org/abs/2512.10899} {arXiv:2512.10899} \BibitemShut
  {NoStop}%
\bibitem [{\citenamefont {Mayers}\ \emph {et~al.}(2018)\citenamefont {Mayers},
  \citenamefont {Tan}, \citenamefont {Egger}, \citenamefont {Rappe},\ and\
  \citenamefont {Reichman}}]{Mayers_2018}%
  \BibitemOpen
  \bibfield  {author} {\bibinfo {author} {\bibfnamefont {M.~Z.}\ \bibnamefont
  {Mayers}}, \bibinfo {author} {\bibfnamefont {L.~Z.}\ \bibnamefont {Tan}},
  \bibinfo {author} {\bibfnamefont {D.~A.}\ \bibnamefont {Egger}}, \bibinfo
  {author} {\bibfnamefont {A.~M.}\ \bibnamefont {Rappe}},\ and\ \bibinfo
  {author} {\bibfnamefont {D.~R.}\ \bibnamefont {Reichman}},\ }\bibfield
  {title} {\bibinfo {title} {How lattice and charge fluctuations control
  carrier dynamics in halide perovskites},\ }\href
  {https://doi.org/10.1021/acs.nanolett.8b04276} {\bibfield  {journal}
  {\bibinfo  {journal} {Nano Letters}\ }\textbf {\bibinfo {volume} {18}},\
  \bibinfo {pages} {8041} (\bibinfo {year} {2018})}\BibitemShut {NoStop}%
\bibitem [{\citenamefont {Nguyen}\ \emph {et~al.}(2025)\citenamefont {Nguyen},
  \citenamefont {Mandal}, \citenamefont {Mahajan},\ and\ \citenamefont
  {Reichman}}]{Nguyen_2025}%
  \BibitemOpen
  \bibfield  {author} {\bibinfo {author} {\bibfnamefont {H.}~\bibnamefont
  {Nguyen}}, \bibinfo {author} {\bibfnamefont {A.}~\bibnamefont {Mandal}},
  \bibinfo {author} {\bibfnamefont {A.}~\bibnamefont {Mahajan}},\ and\ \bibinfo
  {author} {\bibfnamefont {D.~R.}\ \bibnamefont {Reichman}},\ }\bibfield
  {title} {\bibinfo {title} {Mixed quantum-classical methods for polaron
  spectral functions},\ }\href {https://doi.org/10.1063/5.0281529} {\bibfield
  {journal} {\bibinfo  {journal} {The Journal of Chemical Physics}\ }\textbf
  {\bibinfo {volume} {163}},\ \bibinfo {pages} {114105} (\bibinfo {year}
  {2025})}\BibitemShut {NoStop}%
\bibitem [{\citenamefont {Keski-Rahkonen}\ \emph {et~al.}(2024)\citenamefont
  {Keski-Rahkonen}, \citenamefont {Ouyang}, \citenamefont {Yuan}, \citenamefont
  {Graf}, \citenamefont {Aydin},\ and\ \citenamefont {Heller}}]{Aydin_PRL2024}%
  \BibitemOpen
  \bibfield  {author} {\bibinfo {author} {\bibfnamefont {J.}~\bibnamefont
  {Keski-Rahkonen}}, \bibinfo {author} {\bibfnamefont {X.}~\bibnamefont
  {Ouyang}}, \bibinfo {author} {\bibfnamefont {S.}~\bibnamefont {Yuan}},
  \bibinfo {author} {\bibfnamefont {A.~M.}\ \bibnamefont {Graf}}, \bibinfo
  {author} {\bibfnamefont {A.}~\bibnamefont {Aydin}},\ and\ \bibinfo {author}
  {\bibfnamefont {E.~J.}\ \bibnamefont {Heller}},\ }\bibfield  {title}
  {\bibinfo {title} {{Quantum-Acoustical {D}rude Peak Shift}},\ }\href
  {https://doi.org/10.1103/PhysRevLett.132.186303} {\bibfield  {journal}
  {\bibinfo  {journal} {Phys. Rev. Lett.}\ }\textbf {\bibinfo {volume} {132}},\
  \bibinfo {pages} {186303} (\bibinfo {year} {2024})}\BibitemShut {NoStop}%
\bibitem [{\citenamefont {Aydin}\ \emph {et~al.}(2024)\citenamefont {Aydin},
  \citenamefont {Keski-Rahkonen},\ and\ \citenamefont {Heller}}]{Aydin_2024}%
  \BibitemOpen
  \bibfield  {author} {\bibinfo {author} {\bibfnamefont {A.}~\bibnamefont
  {Aydin}}, \bibinfo {author} {\bibfnamefont {J.}~\bibnamefont
  {Keski-Rahkonen}},\ and\ \bibinfo {author} {\bibfnamefont {E.~J.}\
  \bibnamefont {Heller}},\ }\bibfield  {title} {\bibinfo {title} {Quantum
  acoustics unravels {P}lanckian resistivity},\ }\bibfield  {journal} {\bibinfo
   {journal} {Proceedings of the National Academy of Sciences}\ }\textbf
  {\bibinfo {volume} {121}},\ \href {https://doi.org/10.1073/pnas.2404853121}
  {10.1073/pnas.2404853121} (\bibinfo {year} {2024})\BibitemShut {NoStop}%
\bibitem [{\citenamefont {Krotz}\ and\ \citenamefont
  {Tempelaar}(2024)}]{Krotz_2024}%
  \BibitemOpen
  \bibfield  {author} {\bibinfo {author} {\bibfnamefont {A.}~\bibnamefont
  {Krotz}}\ and\ \bibinfo {author} {\bibfnamefont {R.}~\bibnamefont
  {Tempelaar}},\ }\bibfield  {title} {\bibinfo {title} {Mixed
  quantum–classical modeling of exciton–phonon scattering in solids:
  Application to optical linewidths of monolayer {MoS}2},\ }\href
  {https://doi.org/10.1063/5.0218973} {\bibfield  {journal} {\bibinfo
  {journal} {The Journal of Chemical Physics}\ }\textbf {\bibinfo {volume}
  {161}},\ \bibinfo {pages} {044117} (\bibinfo {year} {2024})}\BibitemShut
  {NoStop}%
\bibitem [{\citenamefont {Li}\ \emph {et~al.}(2021)\citenamefont {Li},
  \citenamefont {Ren},\ and\ \citenamefont {Shuai}}]{Li_2021}%
  \BibitemOpen
  \bibfield  {author} {\bibinfo {author} {\bibfnamefont {W.}~\bibnamefont
  {Li}}, \bibinfo {author} {\bibfnamefont {J.}~\bibnamefont {Ren}},\ and\
  \bibinfo {author} {\bibfnamefont {Z.}~\bibnamefont {Shuai}},\ }\bibfield
  {title} {\bibinfo {title} {A general charge transport picture for organic
  semiconductors with nonlocal electron-phonon couplings},\ }\href
  {https://doi.org/10.1038/s41467-021-24520-y} {\bibfield  {journal} {\bibinfo
  {journal} {Nature Communications}\ }\textbf {\bibinfo {volume} {12}},\
  \bibinfo {pages} {4260} (\bibinfo {year} {2021})}\BibitemShut {NoStop}%
\bibitem [{\citenamefont {Tanaskovi\ifmmode~\acute{c}\else \'{c}\fi{}}\ \emph
  {et~al.}()\citenamefont {Tanaskovi\ifmmode~\acute{c}\else \'{c}\fi{}},
  \citenamefont {Makrushin},\ and\ \citenamefont {Mitri\ifmmode~\acute{c}\else
  \'{c}\fi{}}}]{Tanaskovic_Zenodo_QC_multimode}%
  \BibitemOpen
  \bibfield  {author} {\bibinfo {author} {\bibfnamefont {D.}~\bibnamefont
  {Tanaskovi\ifmmode~\acute{c}\else \'{c}\fi{}}}, \bibinfo {author}
  {\bibfnamefont {M.}~\bibnamefont {Makrushin}},\ and\ \bibinfo {author}
  {\bibfnamefont {P.}~\bibnamefont {Mitri\ifmmode~\acute{c}\else \'{c}\fi{}}},\
  }\href {https://doi.org/10.5281/zenodo.19130430} {\bibinfo {title} {Data for
  "{Q}uantum-classical study of charge transport in organic semiconductors with
  multiple low-frequency vibrational modes" [{D}ata set]}},\ \bibinfo {note}
  {{Z}enodo (2026)}\BibitemShut {NoStop}%
\end{thebibliography}

\end{document}